\documentclass[aps,prd,a4paper,twocolumn,amsmath,showpacs,superscriptaddress,nofootinbib,preprintnumbers,longbibliography]{revtex4-1}
\usepackage{graphicx}
\usepackage{epsf}
\usepackage{bm}
\usepackage{amsmath}
\usepackage{amsfonts}
\usepackage{amssymb}
\usepackage{epstopdf}
\usepackage{natbib}
\usepackage[dvipsnames,table]{xcolor}
\usepackage{physics}
\usepackage{hyperref}
\usepackage{lipsum}
\usepackage{float}
\usepackage{multirow}

\usepackage[capitalize]{cleveref}
\providecommand{\U}[1]{\protect\rule{.1in}{.1in}}
\usepackage[normalem]{ulem}
\usepackage[dvipsnames]{xcolor}
\hypersetup{colorlinks,linkcolor={magenta},citecolor={red},urlcolor={blue}}  

\usepackage{titlesec}
\usepackage[T1]{fontenc}
\usepackage{xcolor}
\usepackage{verbatim}

\newcommand{\jav}[1]{\textcolor{red}{(jav: #1)}}

\newcommand{\luis}[1]{\textcolor{blue}{(luis: #1)}}

\begin{document}


\title{Non-parametric reconstruction of cosmological observables using Gaussian Processes Regression.}


\date{\today}

\author{Jos\'e de Jes\'us Vel\'azquez}
\email{jess_2497@ciencias.unam.mx}
\affiliation{Facultad de Ciencias, Universidad Nacional Aut\'{o}noma de M\'{e}xico, Circuito de la Investigaci\'on Cient\'ifica Ciudad Universitaria, CDMX, 04510, Mexico}

\author{Luis A. Escamilla}
\email{luis.escamilla@icf.unam.mx}
\affiliation{Instituto de Ciencias F\'isicas, Universidad Nacional Aut\'onoma de M\'exico, Cuernavaca, Morelos, 62210, M\'exico}
\affiliation{School of Mathematics and Statistics, University of Sheffield, Hounsfield Road, Sheffield S3 7RH, United Kingdom}

\author{Purba Mukherjee}
\email{pdf.pmukherjee@jmi.ac.in}
\affiliation{Centre for Theoretical Physics, Jamia Millia Islamia, New Delhi 110025, India}

\author{J.~Alberto V\'azquez}
\email{javazquez@icf.unam.mx}
\affiliation{Instituto de Ciencias F\'isicas, Universidad Nacional Aut\'onoma de M\'exico, Cuernavaca, Morelos, 62210, M\'exico}

\begin{abstract}


The current accelerated expansion of the Universe remains ones of the most intriguing topics in modern cosmology, driving the search for innovative statistical techniques. Recent advancements in machine learning have significantly enhanced its application across various scientific fields, including physics, and particularly cosmology, where data analysis plays a crucial role in problem-solving. In this work, a non-parametric regression method with Gaussian processes is presented along with several applications to reconstruct some cosmological observables, such as the deceleration parameter and the dark energy equation of state, in order to contribute with some information that helps to clarify the behavior of the Universe. It was found that the results are consistent with $\Lambda$CDM and the predicted value of the Hubble parameter at redshift zero is $H_{0}=68.798\pm 6.340(1\sigma) \text{ km}\text{ s}^{-1}\text{ Mpc}^{-1}$.

\end{abstract}
\maketitle

\section{Introduction}\label{sec:intro}

Nowadays it is becoming more common to hear that we are currently living in the Golden Age of Cosmology, 
whose origin goes back to the early 90's when the Cosmic Background Explorer (COBE) satellite was launched 
in order to provide information of the Cosmic Microwave Background \cite{boggess1992cobe}. 
This event marked the beginning of a series of outstanding discoveries such as the necessity to incorporate 
the Dark Matter (DM) and Dark Energy (DE) components to account for the structure formation and 
the current accelerated expansion of the Universe, which later on gave rise to the standard cosmological model 
or Lambda Cold Dark Matter (LCDM) (more on this model later). 
This Golden Age is also characterized by the huge amount of observations and data obtained as a 
result of several world-wide collaborations, such as Planck \cite{aghanim2020planck}, SDSS \cite{york2000sloan}, 
SNLS \cite{astier2006supernova}, DESI \cite{Dey_2019}, JWST \cite{gardner2023james} and Euclid \cite{laureijs2011euclid}, to mention a few. 
This was definitely a remarkable achievement since it provided the community with valuable information 
to work with, but it also came with a set of obstacles, such as how to process and analyze the avalanche of  
new data. Fortunately, around the same time, a new field of mathematics was starting to grow in strength: 
Machine Learning.
\\

Machine Learning (ML) is the subfield of Artificial Intelligence dedicated to the mathematical modeling 
of data. It is a method to find solutions to problems by using computers, which differs from regular 
programming since the latter takes data and rules to return results. In contrast, ML 
takes data and results to deduce the rules that relate them. A ML system is said to be trained rather 
than programmed \cite{chollet2017deep}. ML can be broadly categorized into three types: supervised, 
unsupervised, and reinforcement learning \cite{theobald2017machine}. 
It can handle a wide variety of problems, 
but the main goal is to learn the process of mapping inputs into outputs, which can then be used 
to predict the outputs for new, unseen, inputs. These algorithms have been widely compared against 
traditional techniques in related fields, obtaining promising results in terms of efficiency and 
performance in favor of ML \cite{sarker2021machine, ray2019quick, qiu2016survey, mahesh2020machine, dhall2020machine}.
The main advantage of ML algorithms is that they can automate repetitive tasks such as data cleaning 
and pattern recognition that might require direct human intervention with traditional methods. 

ML also contains valuable tools for the process known as \textbf{reconstructions}. 
In the absence of a fundamental and/or well-defined theory, a reconstruction is able to analyze a physical quantity and provide some insights of its general behavior; it can be broadly categorized into parametric and non-parametric \cite{sahni2006reconstructing}. 
In parametric reconstructions, the target quantity is studied by proposing a particular function with free parameters that should be inferred using observations. These functions are commonly phenomenological parameterizations to model, for instance, the DE equation of state $w(z)$, 
which could bring some information of the DE's fundamental nature. 
Non-parametric reconstructions, on the other hand, instead of focusing on a particular function they apply numerical or statistical tools directly to the data. This approach allows to decrease the bias towards a particular behavior, however they are prone to overfitting or may produce inaccurate results on extrapolations outside the range spanned by the data. In this work the main focus is in non-parametric reconstructions, where ML algorithms shine through.
That is, ML algorithms allow to predict the behavior of some observable quantities, 
even when an exact model of them is not fully specified \cite{muller2016introduction}.
Some useful and popular supervised learning methods that have been applied to Cosmology are:
\\

\textbf{Artificial Neural Networks (ANN)}: Named so because of their analogy to the behavior of 
the human brain. ANN are made up of layers of sets of units called neurons that individually 
process data inputs. Each neuron is connected to the others through links with weights that 
are evaluated by an activation function, discarding the worst options and prioritizing the best 
ones. ANN are commonly used to solve classification and pattern recognition problems in images, 
speech, or signals. ANN have also predictive applications in the financial \cite{Sewell} 
or atmospheric \cite{ABHISHEK2012311} sector. The field of Cosmology is no stranger to Neural 
Networks, just to mention a few examples we have: CosmicNet I \cite{albers2019cosmicnet} and 
CosmicNet II \cite{gunther2022cosmicnet}, which are used to accelerate Einstein-Boltzmann 
solvers; physically-informed neural networks as a replacement for numerical solvers for 
differential equations in cosmological codes \cite{chantada2023cosmology,chantada2024faster}; 
a more suited application consists on using ANN directly with data to non-parametrically 
reconstruct certain cosmological quantities such as the Hubble parameter and structure formation 
through $f\sigma_8$ \cite{Gomez-Vargas:2021zyl}, deceleration parameter \cite{Mukherjee:2022yyq}, 
rotation curves \cite{Garcia-Arroyo:2024rdj}; on scalar-tensor theories \cite{Dialektopoulos:2023jam}; 
or to test the cosmic distance ladder
\cite{Mukherjee:2024akt,Shah:2024slr}; to emulate functions such as the power spectrum
\cite{Agarwal:2012ew, Agarwal:2013aea, Costanza:2023cgt} or to speed up computational process
\cite{Gomez-Vargas:2024izm, Nygaard:2022wri, Sikder:2022hzk, Jense:2024llt}, along with Genetic Algorithms \cite{Gomez-Vargas:2022bsm};  
for an introduction of ANN in Cosmology, see \cite{Olvera:2021jlq}.
\\

\textbf{Decision Trees and Random Forests (RF)}: Essentially, Decision Trees learn a 
hierarchy of if/else questions and reach an appropriate decision. Decision trees can be used in 
marketing campaigns \cite{DTmarketing} or diagnosis of diseases \cite{DTdiagnosis} to mention a 
few examples. Random Forests are based on a set of Decision Trees that are uncorrelated and merged 
to create more accurate data predictions. These types of algorithms are often used to solve 
classification problems \cite{scikit-learn}, which can be of great use in the field of Cosmology. 
Some examples are: Gravitational Waves' classification \cite{shah2023waves, baker2015multivariate}, 
joint redshift-stellar mass probability distribution functions \cite{mucesh2021machine} and 
N-body simulations \cite{conceiccao2024fast, chacon2022classification, chacon2023analysis}.
\\

\textbf{k-Nearest Neighbors (k-NN)}: This algorithm consists of storing the training dataset 
and formulating a method that finds the closest data values to make predictions for a new test 
data point. It is possibly the simplest ML method and has a wide spectrum of applications, 
such as the creation of customized recommended systems \cite{KNN}. Given the ease with which 
k-NN finds groups/agglomerations, its use in cosmology has focused on topics related to 
structure formation such as galaxy-clustering \cite{banerjee2021nearest, banerjee2022modelling, yuan20232d, wang2022detection}.
\\

There is another ML technique, which works as the basis for this paper and it is known as 
\textbf{Gaussian Process Regression} (GPR). 
{Over the last decade, GPR has become particularly popular in cosmology for testing the 
concordance model \cite{Nair:2015jua,Rana:2017sfr,Mukherjee:2021kcu,Mukherjee:2023yxq}, 
cosmographic studies \cite{Shafieloo_2012,Mukherjee:2020ytg,Mukherjee:2020vkx,Jesus:2022xwb,Dinda:2023xqx}, reconstruction of 
parameters that characterize the cosmic expansion \cite{GPC, Mukherjee:2024ryz, Mukherjee:2023lqr,Calderon:2023msm,LHuillier:2019imn}, 
reconstructing dark energy \cite{Holsclaw_2011, seikel2012reconstruction, Zhang:2018gjb, Dinda:2024ktd, Dinda:2024xla, Calderon:2022cfj}, constraining 
spatial curvature \cite{Yang:2020bpv,Dhawan:2021mel,Mukherjee:2022ujw, Dinda:2022vmb, Dinda:2023svr}, exploring the interaction 
between dark matter and dark energy \cite{Yang:2015tzc,Mukherjee_2021, Cai_2017, Bonilla:2021dql, Escamilla:2023shf}, testing modified theories of gravity\cite{Zhou_2019,Belgacem_2020,Yang_2021,LeviSaid:2021yat,Bernardo:2021qhu,Dialektopoulos:2023jam,Gadbail:2024rpp},testing consistency among datasets 
\cite{Keeley:2020aym}, emulating the matter power spectrum \cite{Ho_2021}, thermodynamic viability analysis
\cite{Banerjee:2023evd, Banerjee:2023rvg}, probing the cosmic reionization history 
\cite{Adak:2024urf, Krishak:2021fxp, Mukherjee:2024cfq} and classification and identification of blended galaxies
\cite{Buchanan_2022}; among many other research fields that take advantage of the ML capabilities 
for analyzing and classifying images, videos, and numerical data. For a pedagogical introduction 
to GPR, one can refer to the Gaussian process website\footnote{\url{http://gaussianprocess.org/}}}. 
Over the course of this work a GPR will be defined and then tested by applying it to the 
prediction of observable quantities in Cosmology. Therefore, the main objective of this work is to provide a basic 
introduction to Gaussian Processes (GPs) and a presentation of some applications of this method through examples. 

The paper is structured as follows: in section~\ref{sec:gp} a general description of the GPR is given; in section~\ref{sec:GPK} we explain the types of existing kernels in the context of a GPR; then in section~\ref{sec:cosmology} we give a brief review of cosmology that yields to the standard cosmological model; in section~\ref{sec:cosmo_gpr} we make use of the GPR methodology on some cosmological quantities and finally in section~\ref{sec:conclusions} we discuss our results and present our conclusions.

\section{Gaussian Processes}\label{sec:gp}

In this section we present some of the relevant concepts before delving into the GPR:
\begin{itemize}
    \item Random Variable 
    is a function that assigns a value to each event in the sample space of a random experiment, 
    it could be either discrete or continuous. 
    For example: when rolling two 6-sided dice the result will be two outcomes $n_1$ and $n_2$. In this case, 
    a discrete random variable $X$ can be the sum of the result of rolling both dice, i.e. $X (n_1,n_2)= n_1+n_2$. 
    In contrast, a continuous random variable could be the weight or height of a population. 
    These quantities, once measured, comprise an interval on the number line, 
    making it continuous when taking an infinite number of possible values.
    
    \item Correlation: also called ``dependence'', it is a statistical relationship between two random variables. 
    For example, when comparing the height of a person with that of their parents, in general, it will be observed 
    that the descendants have heights similar to the progenitors, this means that there is a connection or 
    positive correlation between both heights. In general, the presence of consequences does not imply causality.
    
    \item Probability distribution: a function that assigns to each event, defined on the random variable, 
    the probability that said event occurs. They can be discrete or continuous. A widely used one is 
    the binomial probability distribution (where there are two possible mutually exclusive events): 
    \begin{equation}
        P(X=k)= \frac{n!}{k!(n-k)!}p^k (1-p)^k,
    \end{equation}
    where $k$ is the number of times an event has occurred, $p$ the probability that said event occurs, 
    and $n$ the number of total events.
    
    \item Normal distribution: also called Gaussian distribution, it is a type of continuous probability 
    distribution with the form:
    \begin{equation}
        f(x) = \frac{1}{\sigma \sqrt{2\pi}}e^{-\frac{1}{2}(\frac{x-\mu}{\sigma})^2},
        \label{normal dist}
    \end{equation}
    where $x$ is a random variable, $\mu$ is the mean and $\sigma$ the standard deviation.
    
    \item Random process: also called stochastic process, it is an object made up of several random variables. 
    An example of a stochastic process is the random walker since each step the walker takes is a 
    random variable. The random variables are not necessarily independent of each other, since there 
    may be correlations as in the Markov chains where the next step in the chain depends on the 
    immediately preceding one.
\end{itemize}


Let $x=X(\omega)$ be the value of a random variable $X$ at $\omega$ and $f(x)$ 
its probability distribution. We say that $X$ is normally distributed if $f$ has the form of 
Eq.~(\ref{normal dist}), which is defined only by the mean $\mu$ and variance $\sigma^{2}$, hence it can be denoted as
%
\begin{equation}
    f(x) \sim N(\mu,\sigma^2).
\end{equation}
Therefore, $N(\mu,\sigma^2)$ is said to be the normal (Gaussian) distribution. If we now 
have an arbitrary number of random variables $x_1, ..., x_n$, then the distribution becomes 
a multivariate normal distribution, which can be denoted as:
\begin{equation}
    \Bar{f}=[f(x_1),...,f(x_n)]\sim \Bar{N}(\Bar{\mu},K(x,x')),
\end{equation}
where $\Bar{\mu}=(\mu(x_1),\mu(x_2),...,\mu(x_n))$ is the vector that contains the means of the random variables and 
\begin{equation}
\small
K(x,x^\prime) = 
\begin{pmatrix}
K(x_1,x_1) & K(x_1,x_2) & \cdots & K(x_1,x_n) \\
K(x_2,x_1) & K(x_2,x_2) & \cdots & K(x_2,x_n) \\
\vdots  & \vdots  & \ddots & \vdots  \\
K(x_n,x_1) & K(x_n,x_2) & \cdots & K(x_n,x_n) 
\end{pmatrix},
\label{kernelmatrix}
\end{equation}
is a matrix with the covariances among the variables. Note that each diagonal element is the covariance of a random variable with itself, which equals its variance.

This reasoning can be extended to the case of a continuous random variable $x$ where each value of $x$ is 
a random variable. In this case, the mean vector becomes a function that returns the mean of the Gaussian 
distribution that defines $x$ and the covariance matrix has to be a function that gives the covariance 
between two continuous random variables $x$ and $x'$. This generalization of a normal distribution for 
continuous random variables is known as a \textit{Gaussian Process}. Therefore, a GP is an infinite collection 
of random variables which is defined by a mean function 
$\mu(x)$ and a covariance function $k(x,x')$, also known as the kernel of the process. Usually, the mean 
$\mu(x)$ is taken to be zero for simplicity, but it can be different with analogous calculations.
\\

\begin{figure}[t!]
\centering
    \includegraphics[width=8.3cm, height=6cm]{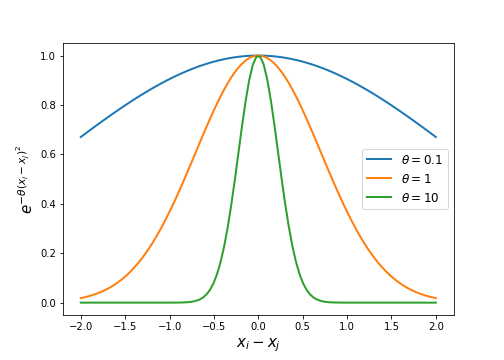}
    \caption{Changes in the correlation when varying the $\theta$ value.}
    \label{fig:theta_valores}
\end{figure}

There are several types of kernels such as the rational quadratic, exponential or Matern (which will be 
further explained in later sections). For example, one of the most commonly used covariance functions 
due to its simplicity and infinite differentiability is the squared exponential kernel, which can be 
written as: 
\begin{equation}
    k(x_i, x_j) = e^{-\theta(x_i-x_j)^2}
    \label{kernel_RBF_theta},
\end{equation}
where the parameter $\theta$ indicates how the correlation is spread, as shown in Figure \ref{fig:theta_valores}. 
The larger the value of $\theta$, the stronger the correlation between variables.

\subsection{Gaussian Process Regression.} \label{sec:GPR}

In order to train a Gaussian Process Regression (GPR) model, a dataset of $n$ points 
$\{(x_{1},y_{1}),(x_{2},y_{2}),...,(x_{n},y_{n})\}$ is needed. Let us define the vectors 
$\vec{x}=(x_{1},x_{2},...,x_{n})$ and $\vec{y}=(y_{1},y_{2},...,y_{n})$. 
The aim of a GPR is to find the posterior probability distribution for the values of the independent 
variable $P(\vec{w}|\vec{y},\vec{x})$, where $\vec{w}$ is a vector of weights that defines the model. 
The posterior is computed by the Bayes' Rule:
\begin{equation}
    P(\vec{w}|\vec{y},\vec{x})=\frac{P(\vec{y}|\vec{x},\vec{w})P(\vec{w})}{P(\vec{y}|\vec{x})}.
    \label{posterior}
\end{equation}
Here: $P(\vec{w})$ is referred to as the \textit{prior} which is a probability distribution that 
contains information about $\vec{w}$ before the observed data; $P(\vec{y}|\vec{x},\vec{w})$ is 
named the \textit{likelihood} and it relates information about the prior distribution with the data; 
the marginal likelihood $P(\vec{y}|\vec{x})$ is a constant of normalization that guarantees the 
posterior is a probability ($0\leq P(\vec{w}|\vec{y},\vec{x}) \leq1$) and it is given by the 
integral of the numerator over all possible values of $\vec{w}$:
\begin{equation}
    P(\vec{y}|\vec{x})=\int{P(\vec{y}|\vec{x},\vec{w})P(\vec{w})}d\vec{w}.
    \label{marginal}
\end{equation}
Note that Bayes' Rule is not restricted to Gaussian distributions, however, in the context of 
GPR, the prior and posterior are both a GP and the data is Gaussian (each value is determined by 
a mean and a standard deviation). For this particular case, the prior and posterior are called 
conjugate distributions with respect to the likelihood function.
\\

The GPR consists in making predictions based on the training data set (also called observables), 
assuming the observations are distributed around a model $f$ with an additive noise $\varepsilon$, 
which is assumed to be Gaussian with zero mean and variance $\sigma^{2}_{n}$:
\begin{equation}
\begin{matrix}
    \vec{y}=f(\vec{x})+\varepsilon,\\
    \text{cov}{(\vec{y})}=K(\vec{x},\vec{x})+\sigma_{n}^{2}I,
    \label{observations}
\end{matrix}
\end{equation}
where $I$ is the identity matrix and $K(\vec{x},\vec{x})$ is the covariance matrix obtained 
when evaluating the kernel in the corresponding training points, that is $[K(\vec{x},\vec{x})]_{ij}=k(x_{i},x_{j})$.

Therefore, it is required to find the test outputs $\vec{f}_{*}$, which are the values of the 
model at the test points $\vec{x}_{*} \equiv \left( {x_1}_*, {x_2}_*, \cdots, {x_n}_* \right)$. 
The posterior distribution of Eq. (\ref{posterior}) can be derived by conditioning the prior 
on the training observations, such that the conditional distribution of $\vec{f}_{*}$ only 
contains those functions from the prior that are consistent with the data set. 
Using the conditioning and marginalizing properties of the Gaussian distribution on the joint distribution for $\vec{f}_{*}$ and $\vec{y}$, it can be proven \cite{Rasmussen} that 
the mean and covariance of the predictions for the test set $\vec{x}_{*}$ is:
\begin{equation}
\begin{matrix}
\bar{\vec{f}}_{*}=\vec{K}_{*}^{\top}(\vec{K}+\sigma_{n}^{2}\vec{I})^{-1}\vec{y} \, , \\
\text{cov}(\vec{f}_{*})=\vec{K}_{**}-\vec{K}_{*}^{\top}\left(\vec{K}+\sigma_{n}^{2}\vec{I}\right)^{-1}\vec{K}_{*} \, .
\label{predictions}
\end{matrix}
\end{equation}
The notation 
$\vec{K}=K(\vec{x},\vec{x})$, $\vec{K}_{*}=K(\vec{x},\vec{x}_{*})$ and $\vec{K}_{**}=K(\vec{x}_{*},\vec{x}_{*})$ is introduced to simplify the 
calculations.


\subsection{Maximum likelihood estimation.} \label{sec:MLE}

Assuming the cases in the training set are independent of each other, the probability density 
of the observations given a set of parameters $\vec{w}$, which is the likelihood from 
Eq. (\ref{posterior}), can be expressed as a product of individual densities
\begin{equation}
    P(\vec{y}|\vec{x},\vec{w})=\prod_{i=1}^{n}p(y_{i}|x_{i},\vec{w}) \, ,
\end{equation}
where $n$ is the number of input training points.
Therefore, using the fact that the product of Gaussian distributions is also Gaussian, 
the marginal likelihood from Eq. (\ref{marginal}), in logarithmic form, becomes the 
\textit{log marginal likelihood}
\begin{equation}
\footnotesize
    \log{P(\vec{y}|\vec{x})}=-\frac{1}{2}\vec{y}^{T}(\vec{K}+\sigma_{n}^{2}\vec{I})^{-1}\vec{y}-\frac{1}{2}\log{|\vec{K}+\sigma_{n}^{2}\vec{I}|}-\frac{n}{2}\log{2\pi}.
    \label{log-marginal}
\end{equation}
Optimal values of the parameters can be estimated by maximizing the log marginal likelihood. This training method used in GPR is known as the \textit{maximum likelihood estimation} \cite{Rasmussen}. The maximizing can be performed by any optimizing algorithm, such as gradient descent or Markov Chain Monte Carlo.

\section{GP Kernel.} \label{sec:GPK}

As seen so far, a fundamental feature of GPR which plays an important role, in the fitting of a model, 
is the kernel. A kernel (or covariance function) describes the covariance (correlation) of the random 
variables of the GP. Together with the mean function, the kernel completely defines a GP. In principle, 
any function that relates two points based on the distances between them can be a kernel, but it must 
satisfy certain conditions in order to represent a covariance function. For a function to be a valid 
kernel, the associated resulting matrix in Eq. (\ref{kernelmatrix}) must be positive definite, which 
implies that it has to be symmetric and invertible.

The covariance function of the variables $x$ and $x'$ is said to be \textit{stationary} if it is a 
function only of {$x-x'$}, since it is invariant under translations, and \textit{non-stationary} otherwise. 
Moreover, if it is a function only of $|x-x'|$ it is \textit{isotropic} since it is invariant under 
rigid transformations. 
\\

As mentioned previously, it is necessary to choose a suitable kernel type for each particular problem. 
The process of creating a kernel from scratch is not always trivial, so it is usual to invoke a 
predefined kernel in order to model a diversity of processes. Some of the most used kernels 
are \cite{scikit-learn}: 
\begin{itemize}
    \item \textbf{Radial Basis Function.}
    \begin{equation}
    k(x,x')=\exp{\left(-\frac{d(x,x')^{2}}{2l^{2}}\right)},
    \label{RBF}
    \end{equation}
    where $d(x,x')$ represents the euclidean distance between $x$ and $x'$ and $l>0$ is known 
    as the length parameter. Sometimes it is written in terms of a value $\theta$ that depends on 
    the length parameter, such as in Eq. (\ref{kernel_RBF_theta}). It is knwon as Radial Basis 
    Function (RBF) because it depends only on the radial distance. 
    Notably, this kernel is \textit{infinitely differentiable}, making it ideal for modeling smooth functions where high regularity is expected.

    \item
    \textbf{Matern.}
    \begin{equation}
    \scriptsize
    k(x,x')=\frac{1}{\Gamma(\nu)2^{\nu-1}}\left(\frac{\sqrt{2\nu}}{l}d(x,x')\right)^{\nu}K_{\nu}\left(\frac{\sqrt{2\nu}}{l}d(x,x')\right),
    \label{Matern}
    \end{equation}
    \normalsize
where $K_{\nu}$ is the modified Bessel function of the second kind, $\Gamma(\nu)$ is the Gamma function, $l$ is the characteristic length scale, and $\nu$ is the smoothness parameter that controls the degree of differentiability of the function. For $\nu = \frac{1}{2}$, the Matern kernel reduces to the \textit{exponential kernel}, which models processes with rough, non-smooth behavior. Notably, for $\nu = 1.5$ and $\nu = 2.5$, the kernel corresponds to once and twice differentiable functions, respectively, allowing for more regular behavior. As $\nu \rightarrow \infty$, the kernel approaches the RBF kernel, which is infinitely differentiable. This flexibility and control over the smoothness makes the Matern kernel especially useful for modeling functions that exhibit varying degrees of smoothness, as often seen in real-world data.
    
    \item \textbf{Exponential Sine Squared (ESS).}
    \begin{equation}
        k(x,x')=\exp{\left(-\frac{2\sin^{2}{(\pi d(x,x')/p)}}{l^{2}}\right)},
        \label{expKernel}
    \end{equation}
where $p>0$ is the periodicity parameter, controlling the periodicity of the kernel, and $l > 0$ is the length scale parameter. This kernel is often called a \textit{periodic kernel} because it models periodic functions with a sinusoidal component, where the periodicity is governed by $p$. The exponential decay modulates the amplitude of the sine function, enabling the kernel to capture periodic behaviors with varying smoothness.

    \item \textbf{Dot Product.}
    \begin{equation}        k(x,x')=\sigma_{0}^{2}+x\cdot x',
    \end{equation}
where $\sigma_{0}$ is a parameter that controls the inhomogeneity or offset of the kernel, while $x \cdot x'$ represents the dot product between the vectors $x$ and $x'$. The term $\sigma_{0}^{2}$ allows for a shift in the kernel's value, providing flexibility in modeling non-zero means or biases in the data.

    \item \textbf{Rational Quadratic (RQ).}
    \begin{equation}
        k(x,x')=\left(1+\frac{d(x,x')^{2}}{2\alpha l^{2}}\right)^{-\alpha},
    \end{equation}
where $\alpha$ is known as the scale mixture parameter, and $l$ is the length scale. The RQ kernel is a scale mixture of \textit{squared exponential kernels}, allowing it to model functions with varying smoothness over different scales. When $\alpha = 1$, the kernel behaves similarly to the RBF kernel. The kernel can model processes with longer-range dependencies for $0< \alpha < 1$, while for $\alpha > 1$, it captures shorter-range correlations.
\end{itemize}
Each of the values that can be varied within the kernel, such as $l$, $\sigma_{0}$, etc. are 
called \textit{hyperparameters}. It is said that GPR is a non-parametric technique because the 
number of hyperparameters is infinite. 
{The reader might have noticed that all kernels described above are stationary (dependent 
on $|x-x'|$), except Dot Product. This dependence on distance alone makes stationary kernels more 
rigid, while also presenting 
poor predictive power when outside the scope of the used data when compared with their non-stationary 
counterparts. Non-stationary kernels are more flexible, which allows for a better estimate outside the scope 
covered by the observations. Nevertheless they are rarely used given the high number of hyperparameters to 
optimize, higher complexity, high computational costs and a greater risk of overfitting when compared 
against stationary ones \cite{paciorek2003nonstationary,paciorek2006spatial,noack2022advanced,noack2024unifying}. 
In this work we will use exclusively stationary kernels and Dot Product, although we 
think that the idea of using 
non-stationary ones for cosmological observations might be worth visiting in a future work. }

Since the kernel is a key feature of GPR, modifying it might produce different models. Therefore, it is 
necessary to establish which kernel is the best option for a particular model. In a real problem, such 
as those presented in Cosmology, the kind of relationship between two variables is not always 
previously known. In these cases, the kernel that results in the best fit after regression may be chosen 
from a set of default kernels. 



\subsection{Kernel selection through $\chi^{2}$.}

A robust statistical tool, known as the $\chi^{2}$ test, could be employed to determine which model, 
derived from various kernels, fits best a specific dataset, thereby enhancing the regression analysis. 
This test evaluates the congruence between two datasets by assessing whether a significant discrepancy exists 
between the observed data values and the model's predictions.

The method consists in defining the objective function $\chi^{2}$ as:
\begin{equation}
    \chi^2= \sum_{i,j}\big(y_i-f(x_i)\big)C_{ij}^{-1}\big(y_j-f(x_j)\big),
    \label{chi2}
\end{equation}
where $(x_{i},y_{i})$ are the data points (or training set), $C_{ij}$ is the covariance matrix and 
$f(x_{i})$ are the values of the model at the independent variable of the data points. When the covariance 
matrix is diagonal we obtain a simplified case for the $\chi^2$ as:
\begin{equation}
    \chi^{2}=\sum_{i}{\frac{[y_{i}-f(x_{i})]^{2}}{\sigma_{y_{i}}^{2}}},
    \label{chi2_2}
\end{equation}
where $\sigma_{y_i}^2$ is the variance and the $i$th element in the diagonal of $C_{ij}$. 
The GPR produces a model data set that can be interpreted as a function $f$ of the independent variable $x$. 
Given an observable $(x_{i},y_{i})$, the numerator of Eq. (\ref{chi2_2}) represents the squared 
distance between the observable and the model for the same value of $x_{i}$. By computing this 
difference over all the available observations (and as such calculating the $\chi^2$ function) we 
can get an idea on how well model $f$ fits the data. 

If the value of $\chi^{2}$ is obtained for models built with different kernels, the best fit will be 
the one that minimizes this objective function. Notice that this method is different from the 
maximum likelihood estimation explained in Section \ref{sec:MLE}, since it is not used to determine the 
hyperparameters as in the training. In this case, the models of regression have been determined 
previously for different kernels and tested to find the best model in terms of the covariance function.
\subsection{A generic example.}

In this section, regression models based on Gaussian Processes are constructed from a 
mock dataset exhibiting a straight-line behavior.
Fortunately, nowadays there is a broad range of standard developed code and libraries that
facilitate performing a Gaussian Process Regression, such as \texttt{GPy} \cite{gpy2014},
\texttt{GPflow} \cite{GPflow}, \texttt{GPyTorch} \cite{gpytorch}, \texttt{PyMC} \cite{PyMC}, 
\texttt{scikit-learn} \cite{scikit-learn} and \texttt{GaPP} \cite{seikel2012reconstruction}. The latter two are 
the ones used during the course of this example and the complete step-by-step procedure can 
be found at the public repository \cite{GPCosmology}. To further simplify, the construction 
of a GPR model consists of 3 steps: 1) specify the prior distribution via the kernel, 2) 
find the hyperparameters that maximize Eq. (\ref{log-marginal}) and 3) evaluate predictions 
with Eqs. (\ref{predictions}) using the optimal hyperparameters and observables.

We will use the function \texttt{GaussianProcessRegressor()}, which initializes a GP prior 
for regression with a specified kernel and its parameters. The method \texttt{fit()} 
returns the same  \texttt{GaussianProcessRegressor()} object fitted to the observables 
using the maximum likelihood estimation. 
This method takes two lists as parameters that 
correspond to the observational data variables $\vec{x}$ and $\vec{y}$. 
Finally, the \texttt{predict()} 
method returns the means and standard deviations of the predictions using Eqs.~(\ref{predictions}).
\\

In the first of our examples of regression, the variances $\sigma_{y_{i}}^{2}$ or noises of 
the observational data are ignored. This approach assumes that the data measurements are 
exact, therefore implying there are no uncertainties or error bars associated to them. 

A mock data set scattered around a linear equation $Y=mX+b$ with $m=3$ and $b=-4$ is created by 
adding a random value between $-15$ and $15$ to $10$ evaluations of the equation at different 
values of $X\in[0,10]$. The aim of the GPR is to reproduce the graph of the line that 
originated the set.

In this case, the $\chi^{2}$ test cannot be used to find an optimal kernel, since 
Eq. (\ref{chi2}) is undefined, thus an alternative method to determine the kernel must be used. 
The predictions of the model using a specific kernel at different values of $X$ will be 
compared via the sum of squared euclidean distances to the points of the original linear 
relationship at the corresponding $X$-values scaled by the number of data points, $n$. 
The result is called Mean Squared Error (MSE) and can be written as:
\begin{equation}
    {\rm MSE}=\frac{1}{n}\sum_{i}{[y_{i}-f(x_{i})]^{2}}.
    \label{MSE}
\end{equation}

The regression model that minimizes the MSE is the one that most resembles the desired line. 
At this point, without loss of generality, the kernel used in this GPR is the Matern  (Eq. (\ref{Matern})).

\begin{figure}[t!]
\centering
    \includegraphics[width=9cm, height=5.5cm]{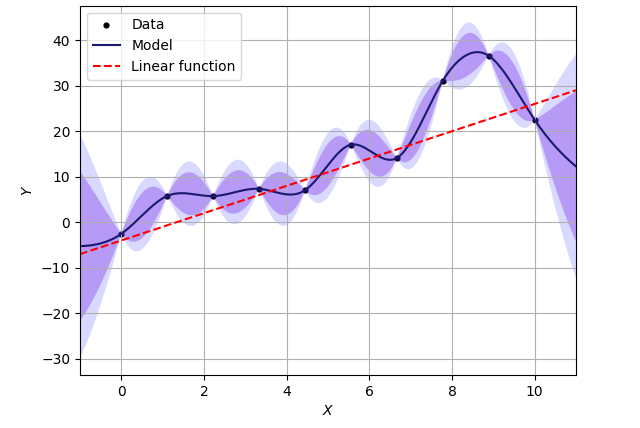}
    \caption{Linear model with null variances in the data. A Matern kernel was used for the reconstruction. It is evident that there is an overestimation of the confidence zone because our data lacks errors.}
    \label{LinealGP}
\end{figure}

Figure \ref{LinealGP} shows the observational mock dataset (black points), the model 
predictions (blue solid line), the line from which the data was obtained (red dash) 
and the confidence zones (in lilac colors) that correspond to $2\sigma$ and $3\sigma$, 
respectively. These confidence intervals will be used for all the regression models 
along this work.

\begin{figure}[t!]
    \centering
    \includegraphics[width=9cm,height=5.5cm]{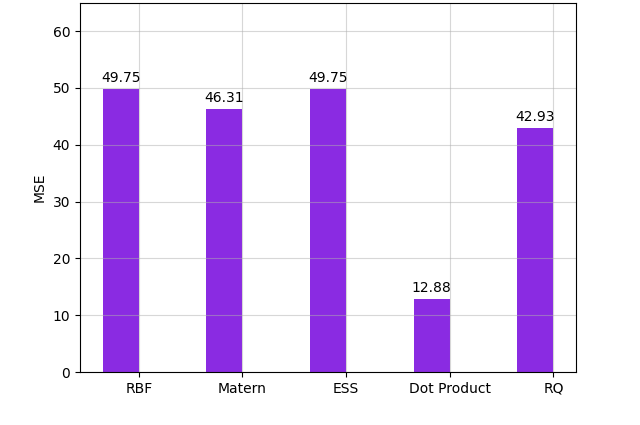}
    \caption{Comparison of the Mean Squared Errors, calculated via Eq. (\ref{MSE}), for 
    different kernels. 
    As can be seen, the model that minimizes the MSE corresponds to the Dot Product kernel, 
    which produces a linear regression.}
    \label{MSEkernels}
\end{figure}

We tested different kernels and the results achieved are quite similar, however, 
as can be seen in Figure \ref{MSEkernels}, the one that minimizes the MSE is the Dot Product kernel.  
The Gaussian process regression for each kernel are 
shown in Figure \ref{KernelsNullLineal}.
The Dot Product kernel produces a linear regression model,  so it is usually 
the best choice when fitting a straight line. In contrast, for the rest of the kernels, the 
uncertainty reduces to zero when the model is evaluated at the observations. This can be 
interpreted as the model overfitting the data, which is expected given that the mock data 
presents no variances \cite{OColgain:2021pyh}. {To mitigate overfitting, one approach is to 
introduce an additional hyperparameter, $\sigma_n$, for noise modeling. This hyperparameter 
accounts for the observational noise, preventing the model from fitting the data too precisely. 
However, adding $\sigma_n$ increases the model complexity, requiring careful tuning to 
balance the bias and variance \cite{Rasmussen}.} \\

\begin{figure}[b!]
\centering
    \includegraphics[width=9cm, height=5.5cm]{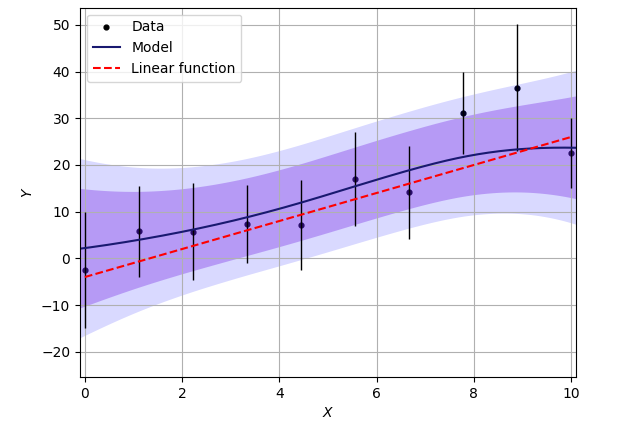}
    \caption{Linear model with variances in the data. An almost linear behavior is observed 
    as we no longer have overfitting. }
    \label{LinealErrorGP}
\end{figure}

On the other hand,
when the observables do have uncertainties (which is the case that most closely resembles 
real data), the variances must be added to the diagonal of the kernel matrix as shown in 
Eq. (\ref{observations}). If these uncertainties come in the form of a non-diagonal covariance 
matrix then it is also added to the kernel so that:
\begin{equation}
    \text{cov}{(\vec{y})}=\vec{K} + \vec{C},
    \label{observations_2}
\end{equation}
with $\vec{C}$ being the covariance matrix of the data. The \texttt{GaussianProcessRegressor()} 
function is able to get as an input an array \texttt{alpha} whose size is equal to the number 
of data that corresponds to the variances associated with each observation. 
The outcome of this approach, illustrated in Figure \ref{LinealErrorGP}, demonstrates that the prediction more closely resembles a straight line, especially when compared to the scenario with a mock dataset with null variances. 

\begin{figure}[h!]
\centering
    \includegraphics[width=8.3cm, height=5.5cm]{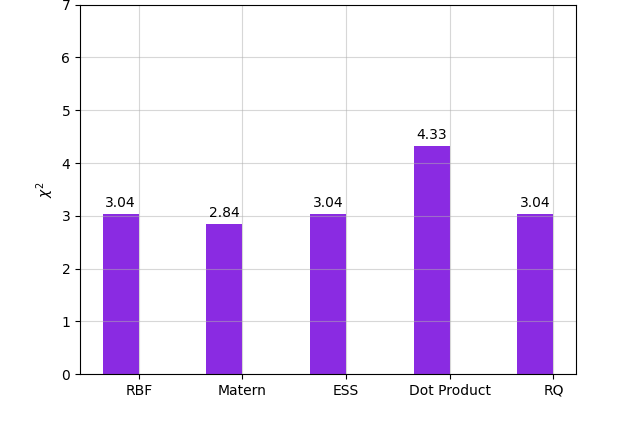}
    \caption{ 
    Different models with a specific kernel are produced to compute 
    their corresponding $\chi^{2}$ value. The kernel with the lowest $\chi^{2}$ generates
    the optimal model, in this case, the Matern kernel.}
    \label{chi2_lineal}
\end{figure}

In this scenario, similar to the MSE test done in Figure~\ref{MSEkernels}, a $\chi^{2}$ test can be employed to determine the kernel that generates 
the optimal model. This involves creating a model for each kernel test, computing the 
$\chi^{2}$ value for each model, and selecting the one with the lowest $\chi^{2}$. 
In Figure \ref{chi2_lineal} we plot the results of this test and, by analyzing it, it can be 
concluded that the model that yields the best fit to our data is the one utilizing a Matern 
kernel, as it produces the lowest value of the objective function. It is crucial to note 
that the model with the lowest $\chi^2$ is not necessarily the best one, as excessively 
minimizing $\chi^2$ can lead to overfitting.
The linear regression models for each kernel are shown in Figure \ref{KernelsNoNullLinear}

\subsection{Derivatives of a GP.}

The RBF kernel (Eq. (\ref{RBF})) is infinitely differentiable and the derivative of a GP is 
also a GP, which allows to reconstruct the derivatives of a function from data. In order to 
reconstruct the derivative, not only the covariance between the observational data is 
required but also the covariance between the function and its derivative and among the 
derivatives of the reconstruction. All of them can be calculated from the derivative of the 
kernel function as described in \cite{seikel2012reconstruction}.

As in Section \ref{sec:GPR}, it can be proven that the mean and covariance of the prediction for the first derivative of this function at test points $\vec{x}_{*} \equiv \left( {x_1}_*, {x_2}_*, \cdots, {x_n}_* \right) $ using a differentiable kernel $k(x_i, x_j)$ are:
\begin{equation}
\begin{matrix}
\bar{\vec{f}^{\prime}_{*}}=\vec{K}_{*}'^{\top}(\vec{K}+\sigma_{n}^{2}\vec{I})^{-1}\vec{y} \, , \\
\text{cov}(\vec{f}^{\prime}_{*})=\vec{K}''_{**}-\vec{K}_{*}'^{\top}\left(\vec{K}+\sigma_{n}^{2}\vec{I}\right)^{-1}\vec{K}_{*}' \, .
\label{predictions_devs}
\end{matrix}
\end{equation}

Here, $\vec{K}_{*}'=K'(\vec{x},\vec{x}_{*}) = \frac{\partial k(x_i, {x_j}_{*})}{\partial {x_j}_{*}}$ and $\vec{K}''_{**}=K''(\vec{x}_{*},\vec{x}_{*}) = \frac{\partial^2 k({x_i}_{*}, {x_j}_{*})}{\partial {x_i}_{*} \partial {x_j}_{*}}$ are introduced to simplify the notation.

\begin{figure*}[t!]
    \begin{centering}\includegraphics[width=15cm, height=9cm]{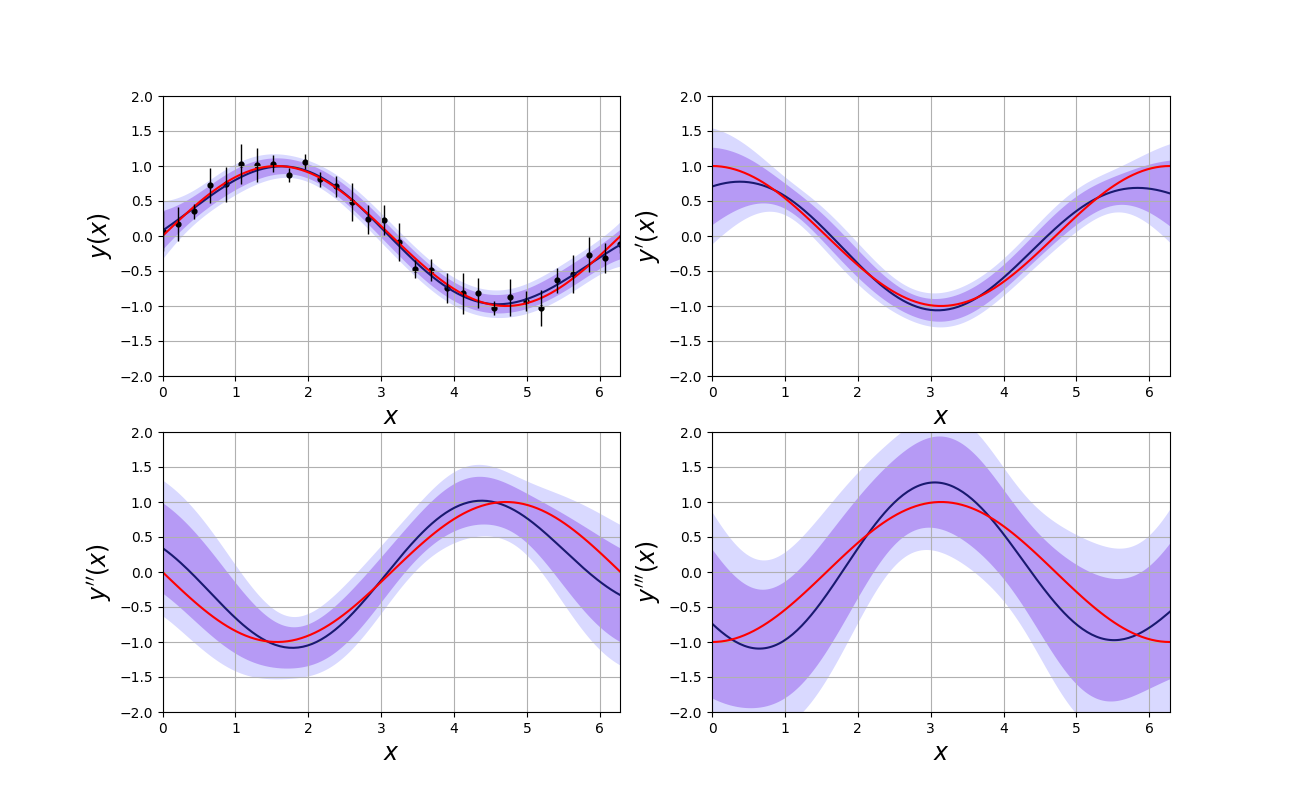}
    \caption{Reconstruction of an example test function $f(x)=\sin{x}$ and its derivatives 
    on $[0,2\pi]$ from a mock data set. The red lines represent the analytical function or 
    derivative and the blue lines are the predictions.
    }
    \label{Devs}
    \end{centering}
\end{figure*}
As can be inferred from these equations, the derivative of the kernel must exist in 
order to compute the predictions of a derivative using GPR. Therefore, an infinitely 
differentiable covariance function is useful when reconstructing a derivative from data, 
this is why an RBF kernel (Eq. (\ref{RBF})) is preferred among others in this type of problems.
If an RBF kernel is used, the procedure can be generalized to any derivative of the model and, 
in particular, the package \texttt{GaPP} \cite{seikel2012reconstruction} allows to compute up to the third 
derivative of a function quickly.
\\

In order to verify the reliability of the code, a mock data set of values scattered around 
a sinusoidal function $y(x)=\sin(x)$ was created by adding a random value between $-0.15$ 
and $0.15$ for different values of $x$. Then the standard deviation (the error bar) of each 
data point was emulated by a random number between $0.1$ and $0.3$. The reconstructions of 
the function and its derivatives are shown in Figure \ref{Devs}. The red lines represent the 
analytical function (the sine function or its derivatives as appropriate) and the blue lines 
are the regression models. The confidence zones correspond to the intervals delimited by 
$2\sigma$ ($95\%$) and $3\sigma$ ($99\%$), where $\sigma$ are the standard deviations of 
the predictions. Note that the analytical function is in the $2\sigma$ interval for all 
the cases, which indicates that the regression is considerably accurate.

Figure $\ref{Devs}$ shows only the scatter plot of the mock data set because the observables 
for the derivatives do not exist. This is an advantage of the technique, since it is 
possible to find the $n$th-derivative of a function only from data values of such function.

\section{Cosmology}\label{sec:cosmology}

Let us start by considering the Universe being homogeneous on scales larger than $150\text{ Mpc}$, 
which means that the distribution of its components does not depend on the position of the observer, 
despite the fact that at short distances the density of matter is perceived as random. 
Likewise, let us also assume the Universe is isotropic, which implies that its properties 
are the same regardless of the direction from which they are observed. The assumption of 
these two characteristics (homogeneity and isotropy at large scales) is known as the 
\textbf{Cosmological Principle} 
and it has been adopted to set restrictions on a great variety of alternative cosmological theories.
    
It is firmly established by observations that our Universe expands \cite{sharov1993edwin}. 
The standard Big Bang model proposes that the Universe emerged about $15$ billion years ago and 
it has been expanding and cooling since then. Measurements using Type IA supernovae as standard 
candles have proven that the expansion of the Universe is also accelerating \cite{perlmutter1999measurements, riess1998observational} and such acceleration is only possible if a substantial fraction of the 
total energy density is a kind of energy with a negative pressure \cite{copeland2006dynamics}. 
This energy component is referred to as \textbf{Dark Energy} (DE) given its unknown nature and origin.
Furthermore, along with DE, another key component, known as \textbf{Dark Matter} (DM), is necessary 
to explain observations regarding structure formation. Given the enigmatic nature of both 
DE and DM, predicting the Universe's long-term future remains an elusive task. Consequently, 
DE and DM represent two of the most compelling and complex challenges in contemporary cosmology.

The expansion of the Universe is described by the Friedmann equations, obtained as solutions 
of the Einstein field equations for the Friedman-Lemaitre-Robertson-Walker (FLRW) metric and 
a perfect fluid with density $\rho$ and pressure $p$. The equations in standard form are:
\begin{eqnarray}
H^2=\left(\frac{\dot{a}}{a}\right)^2=\frac{8\pi G}{3}\rho-\frac{kc^2}{a^2}, \nonumber \\
        \frac{\ddot{a}}{a}=-\frac{4\pi G}{3}\left(\rho+\frac{3p}{c^{2}}\right),
\label{friedmann_eqs}
\end{eqnarray}
where $a$ is known as the scale factor, which is a dimensionless function of time and is related 
to the size of the Universe; $\dot{a}$ and $\ddot{a}$ denote the first and second derivative 
of $a$ with respect to the cosmic time; $H$ is the Hubble parameter, which describes how fast 
the Universe is expanding; 
$G$ is the gravitational constant; 
$c\approx3\times10^{5}\text{ km}/\text{s}$ is the speed of light in vacuum and $k$ is the 
curvature parameter, which determines the shape of the Universe \cite{mukhanov}.\\

One of the most favored models by evidence is the $\Lambda$CDM. This model proposes that the DM 
component of the Universe is a non-relativistic (cold) that only interacts gravitationally, while the DE is due 
to an unknown component represented by the cosmological constant $\Lambda$. 
As mentioned previously, DE is an exotic component in the energy budget of the Universe, which 
is theorized to be responsible for its accelerated expansion. 
Most cosmological models consider DE to be a perfect fluid, which means that it is 
incompressible and with zero viscosity. Then it follows that, for a perfect fluid, its equation 
of state (EoS) is characterized by a dimensionless value $w$. In the case of barotropic fluids $w$ 
given by the proportionality function between its pressure $p$ and energy density~$\rho$:
$$
{p}=c^2w{\rho}.
$$
For perfect fluids such as baryonic matter and relativistic matter (radiation) their EoS's 
are $w=0$ and $w=1/3$, respectively.   Understanding the behavior of the Dark Energy's 
equation of state is a focal point of contemporary cosmology. It is established that the pressure 
exerted by DE must be negative, given its role in driving cosmic expansion instead of contraction. 
Furthermore, accelerated expansion is predicted to occur when the equation of state parameter falls below $-1/3$. When working with the $\Lambda$CDM model one assumes that $w=-1$ for the DE, giving its characteristic behavior of a cosmological constant.

For the standard cosmological model, taking into consideration the equations of state for 
every component when solving Eq. \ref{friedmann_eqs}, the Hubble parameter obtained from the 
first Friedmann equation in terms of the present values of the density parameters $\Omega_i$ is:
\begin{equation}
    \small{H(z)=H_{0}\sqrt{\Omega_{r,0}a^{-4}+\Omega_{m,0}a^{-3}+\Omega_{k,0}a^{-2}+\Omega_{\Lambda,0}}}, \,
    \label{Hfriedmann}
\end{equation}
where the density parameters are $\Omega_{r,0}$ for radiation, $\Omega_{m,0}$ for the matter sector, 
which includes DM and baryons, $\Omega_{k,0}$ to account for the spatial curvature, 
$\Omega_{\Lambda,0}$ to describe the vacuum density in the form of a cosmological constant 
(this represents the DE component) and $H_{0}$ the Hubble parameter, known as the Hubble constant. 
The subscript ``0'' means that they are evaluated at the present time. 
For a spatially flat model ($\Omega_{k}=k=0$) we have $\Omega_{m}+\Omega_{r}+\Omega_{\Lambda}=1$. 



In order to determine a concept of distance between two objects in the Universe, it is 
convenient to present some common definitions of distance measures in 
Cosmology \cite{peebles, Hogg:1999ad}, these include:
\\

\noindent \textbf{1. Comoving distance}: Due to the homogeneity of the Universe, it is 
possible to define a coordinate system that considers the expansion of the Universe. The distance 
between two objects in this system remains constant, so the comoving distance is defined 
as 
    \begin{equation}
        d_{C}(z)=d_{H}\int_{0}^{z}{dz'\frac{H_{0}}{H(z')}},
\end{equation}
where $d_{H}=\frac{c}{H_{0}}$ is the Hubble distance. \\

\noindent    \textbf{2. Transverse comoving distance}: When considering the curvature 
intrinsic to the geometry of space-time, expressed by the parameter $\Omega_k$, the 
transversal comoving distance is defined as,
    \begin{equation}
        d_{M}=\begin{cases}\frac{d_{H}}{\sqrt{\Omega_{k}}}\sinh{\left(\frac{\sqrt{\Omega_{k}}d_{C}(z)}{d_{H}}\right)}&\text{ if }\Omega_{k}>0,\\
        d_{C}(z)&\text{ if }\Omega_{k}=0,\\
        \frac{d_{H}}{\sqrt{-\Omega_{k}}}\sin{\left(\frac{\sqrt{-\Omega_{k}}d_{C}(z)}{d_{H}}\right)}&\text{ if }\Omega_{k}<0,
        \end{cases}
    \end{equation}
which is equal to the comoving distance in the case of a flat space-time, i.e. for $\Omega_k=0$. \\

\noindent    \textbf{3. Luminosity distance}: Comparing the absolute and apparent magnitudes 
between two objects, that is, the actual brightness emitted by an object compared to the 
brightness observed from Earth, the luminosity distance is defined, which is written in 
terms of the transverse comoving distance as:
    \begin{equation}
        d_{L}(z)=(1+z)d_{M}(z).
    \end{equation}

From the above equations, the normalized comoving distance is also obtained as,
\begin{equation}
    D(z)= \frac{1}{d_{H}} \left(\frac{1}{1+z}\right)d_{L}(z).
\end{equation}
In the particular case of a flat Universe, a simple expression for the derivative of the 
normalized comoving distance can be obtained:
\begin{equation}
    D'(z)=\frac{H_{0}}{H(z)}.
    \label{Dprime}
\end{equation}

\noindent {The cosmological quantities are broadly categorized into two groups - 
the physical (dynamical) quantities like the DE equation of state parameter $w$, vs the 
kinematical (cosmographical) quantities that are defined as time derivatives of the scale 
factor $a$, for example, the Hubble $H$, deceleration $q$ and jerk 
 $j$ parameters. The deceleration parameter is defined as:
\begin{equation}
    q=-\frac{\ddot{a}a}{\dot{a}^{2}},
\end{equation}
which can be written in terms of the derivatives of $D(z)$ with respect to the redshift $z$, 
as
\begin{equation}
    q(z)=-1 -\frac{D''(z)}{D'(z)}(1+z),
    \label{q}
\end{equation}
or, in terms of $H(z)$ and its derivative $H'(z)$, as
\begin{equation}
    q(z)=-1-\frac{H'(z)}{H(z)}(1+z) \, .
\end{equation}
The deceleration parameter is a measure of the acceleration of the expansion of space, 
and it is said to be accelerating when $q$ becomes negative~\cite{peebles}.

Furthermore, with DE having a time-varying dynamical equation of state $w(z)$ (ignoring the 
contribution from spatial curvature and radiation), we can write the Hubble parameter $H(z)$ 
by integrating the Friedmann equation \eqref{friedmann_eqs} as,
\begin{equation}
\frac{H^2(z)}{H_0^2} =\Omega_{m,0}(1 + z)^3 + (1 - \Omega_{m,0}) e^{3 \int_0^z \frac{1 + w(x)}{1 + x} dx} . 
\end{equation}
On differentiating the above equation one can arrive at this expression for the DE equation of 
state $w(z)$, as
\begin{equation}\label{eq:w_z}
w(z) = \frac{2(1 + z)H(z) H'(z) - 3H^2(z)}{3H^2(z) - \Omega_{m,0} H_0^2(1 + z)^3} \, .
\end{equation}

As the deceleration parameter $q$ is now estimated and found to be evolving, we focus 
on the next higher-order derivative, the jerk parameter $j$, defined as
\begin{equation}
    j = \frac{\dddot{a}}{aH^3} \, .
\end{equation}  
It can be rewritten as a function of redshift $z$, in terms of the Hubble parameter $H$ 
along with its derivatives $H'(z)$ and $H''(z)$, as
\begin{equation} \label{eq:j}
    j(z) = 1 - 2(1 + z) \frac{H'}{H} + (1 + z)^2 \left( \frac{H''}{H} + \frac{H'^2}{H^2} \right) \, .
\end{equation}
For the $\Lambda$CDM model $j$ is exactly unity. So, any non-monotonic evolution of $j$ can help in understanding the nature of dark energy in the absence of any convincing physical theory \cite{Alam:2003sc, Sahni:2002fz}.
}

\begin{figure}[t!]
\centering
    \includegraphics[width=8cm, height=5.5cm]{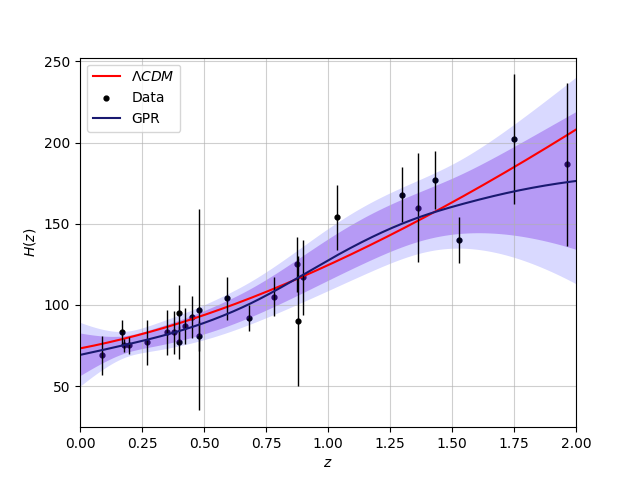}
    \caption{Hubble parameter reconstruction model using a Matern kernel.}
    \label{ModeloHubble}
\end{figure}

\section{Cosmological functions with GPR}\label{sec:cosmo_gpr}
\medskip

\subsection{Hubble parameter}

In Cosmology, the aim is to find a mathematical description that explains the characteristics 
of the Universe and predicts its evolution over time. Thus determining the dependency of $H$ 
as a function of $z$ is one of the main topics of study in Cosmology. \\

For the above, the regression method with GP is a very useful tool as it allows to 
reconstruct the evolutionary model from certain observational data. In this case, the data 
will be Hubble parameter observations for different redshifts from cosmic chronometers as an 
alternative to the commonly used data from Type Ia Supernovae. There is a set of $31$ data 
points for $H(z)$ obtained by different authors, which have been gathered and used in many 
works, such as \cite{G_mez_Valent_2018} and \cite{Vagnozzi_2021}. 
Using the developed code that contains variances and the Hubble parameter data, the model 
shown in Figure~\ref{ModeloHubble} is obtained. 

The curve for $H(z)$ in the $\Lambda$CDM model was created from Eq.~(\ref{Hfriedmann}) and 
the values for the density parameters given by Planck results \cite{aghanim2020planck}, 
were obtained under the assumption of a flat Universe as $\Lambda$CDM.

Various models were tested with different kernels as shown in Figure \ref{chi2bykernel}. 
The model that minimizes $\chi^{2}$ for $H(z)$ was produced by a Matern kernel with the 
default initial characteristic length of $l=1$ and an order of $\nu=1.5$. The optimized 
hyperparameter after the training is $l=4.1$. One key result to stress out
is that from the family of kernels the least preferred corresponds to the Dot Product, that is, 
the equation of state for the dark energy is incline to be anything else expect a linear regression, in particular, a constant equation of state.
\\

Evaluating the model for $z=0$, the value for the Hubble constant $H(0)=H_0=68.79\pm 6.34(1\sigma) 
\text{ km}\text{ Mpc}^{-1} \text{ s}^{-1}$ is obtained as can be seen at~\cite{GPCosmology}. 
While the current analysis focuses solely on the cosmic chronometer Hubble data, 
it is worth noting that incorporating other datasets, such as the Type Ia Supernovae (SNIa) \cite{Scolnic:2021amr, Brout:2022vxf} or 
Baryon Acoustic Oscillations (BAO) \cite{eBOSS:2020yzd, DESI:2024mwx, DESI:2024uvr}, could provide a more comprehensive approach to determining 
$H_0$. However, unlike the cosmic chronometers, both these datasets do not offer independent measurements 
of $H_0$. For SNIa, astrophysical modeling of the absolute magnitude $M_B$ is required, and for BAO, 
the sound horizon $r_d$ must be known. This calls for additional modeling to properly handle the GP 
kernel hyperparameters and cosmological parameters in a coherent manner \cite{Hwang:2022hla}, 
which is beyond the scope of the present work. However, we plan to expand the analysis by incorporating these 
additional datasets, which could improve the precision of $H_0$ estimates \cite{Haridasu:2018gqm, G_mez_Valent_2018, Mukherjee_2021, LHuillier:2019imn, Dinda:2022jih} 
and offer a more robust comparison in future studies.

\subsection{Dark Energy equation of state}

\begin{figure}[t!]
\centering
    \includegraphics[width=8cm, height=5.5cm]{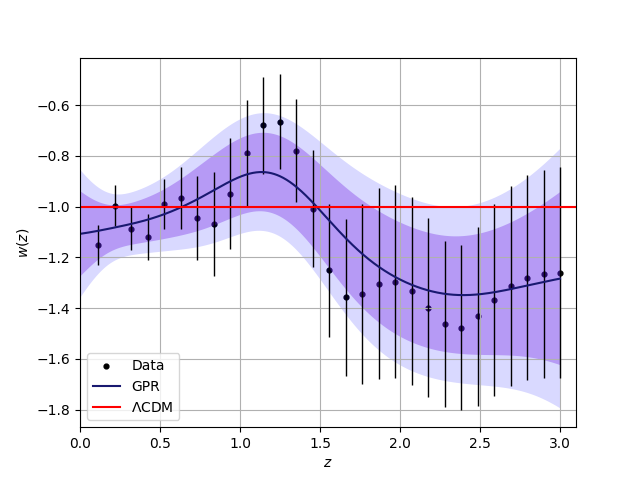}
    \caption{Reconstruction model for the dark energy equation of state, using a
     mock dataset.
    }
    \label{ModeloEO}
\end{figure}

\begin{figure}[t!]
\centering
    \includegraphics[width=8cm, height=5.5cm]{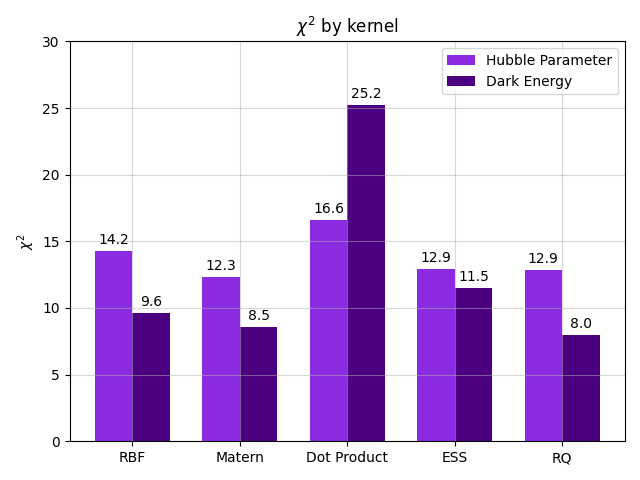}
    \caption{
    Models with different kernels were produced for both the Hubble Parameter and the 
    Dark Energy functions, and computed its $\chi^{2}$ values. As observed, the optimal covariance 
    function was the Matern for Hubble Parameter and the Rational Quadratic for the Dark Energy.
    }
    \label{chi2bykernel}
\end{figure}

\begin{figure}
    \centering
    \includegraphics[width=8cm, height=5.5cm]{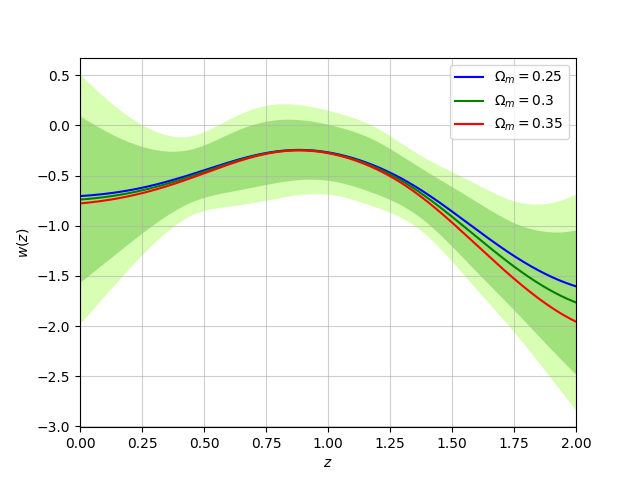}
    \caption{Reconstruction model for the dark energy equation of state, from the
    reconstructing of $H(z)$ using 
    from Cosmic Chronometers, through Eq.~\ref{eq:w_z}.}
    \label{w_z_CC}
\end{figure}

If the DE is considered as a dynamic component, then its EoS should be different from $-1$ 
(so as to differentiate itself from $\Lambda$CDM), or it could present a dependence on redshift 
as $w(z)$. As a proof of the concept and using the previously established methods, we will 
use a mock dataset, which contains information of the BAO and SNIa datasets\footnote{The data points used here come from a model-independent reconstruction 
of the DE EoS from \cite{escamilla2023model}. 
} from the dark energy equation of state as a function of $z$ to reconstruct it. As such, 
a non-parametric model of $w(z)$ with GPR using a RQ kernel will be obtained 
(Figure $\ref{ModeloEO}$). In Figure \ref{chi2bykernel}, a comparison of the values of $\chi^{2}$ 
for models obtained using different kernels is shown and the different model can be seen 
in Figure \ref{DEkernels}.
Note that when reconstructing $H(z)$, the RQ kernel was the best option since it returned 
the minimum values. As already stated earlier, the $\Lambda$CDM model EoS for DE is proposed 
as a constant of value $-1$, so if there are cracks in the standard model then our 
reconstruction should find deviations from this value. In our case, it was found, from the
mock dataset, that  $w=-1$ is well within 1$\sigma$ bounds of the reconstruction using GPR, which 
can be interpreted as a minor evidence against $\Lambda$CDM. 
For the sake of comparison, we perform the reconstruction of $w(z)$ through
Eq.~\ref{eq:w_z}, but now by using the cosmic chronometers dataset. 
In general terms, the results obtained in Figure~\ref{w_z_CC} resembles
the one presented in Figure~\ref{ModeloEO}, with a maximum value around $z\sim 1$ and the crossing
of the phantom divide line ($w=-1$) at about $z\sim 1.5$. Error bars correspond to $\Omega_{m,0}=0.3$, and to examine the sensitivity of the reconstruction, we have included two more values of $\Omega_{m,0}$.

\subsection{Reconstruction of the deceleration parameter \label{sec:q}}

\begin{figure}
    \centering
    \includegraphics[width=8cm, height=5.5cm]{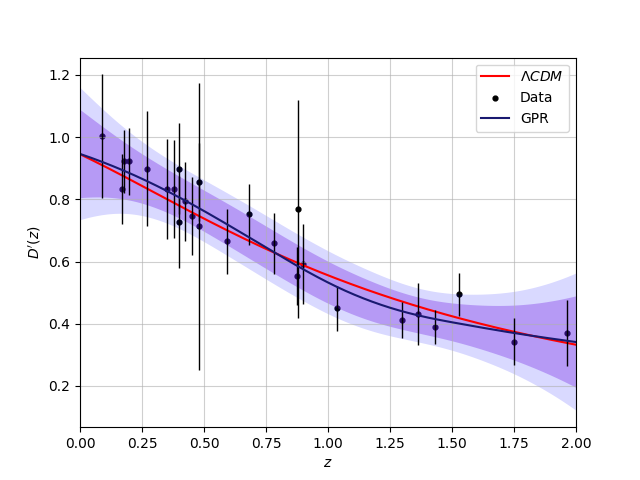}
    \caption{Reconstruction of the first derivative of the normalized comoving distance.}
    \label{Dp}
\end{figure}

By using $H(z)$ data from cosmic chronometers, the predicted value of 
$H_0=68.798 \text{ km}\text{ Mpc}^{-1} \text{ s}^{-1}$ 
and Eq.~(\ref{Dprime}), we can obtain a derived dataset of $D'(z)$. To obtain the 
variances/errors of this new dataset it is straightforward to use the approximation 
of ratio distribution for uncorrelated variables\footnote{The variance of a ratio 
distribution $\frac{X}{Y}$ of two uncorrelated random variables $X$ and $Y$ can be 
approximated with a Taylor expansion around $\mu_{X}$ and $\mu_{Y}$ as \cite{kendall}: 
$\text{Var}\left(\frac{X}{Y}\right)=\frac{\mu_{X}^{2}}{\mu_{Y}^{2}}\left[\frac{\text{Var}(X)^{2}}{\mu_{X}^{2}}+\frac{\text{Var}(Y)^{2}}{\mu_{Y}^{2}}\right]$}. So far, the Matern kernel has presented 
the most suitable models (at least regarding the $\chi^2$ obtained), henceforth this kernel 
will be used for the reconstruction. The resulting GPR prediction for $D'(z)$ from the 
derived dataset and a comparison with the $\Lambda$CDM values computed by combining 
Eq. (\ref{Dprime}) and Eq. (\ref{Hfriedmann}) with the corresponding density parameters and 
the value of $H_{0}=67.32 \text{ km}\text{ Mpc}^{-1} \text{ s}^{-1}$ from Planck 
results \cite{aghanim2020planck} are shown in Figure \ref{Dp}.

\begin{figure}[t!]
    \includegraphics[width=8cm, height=5.5cm]{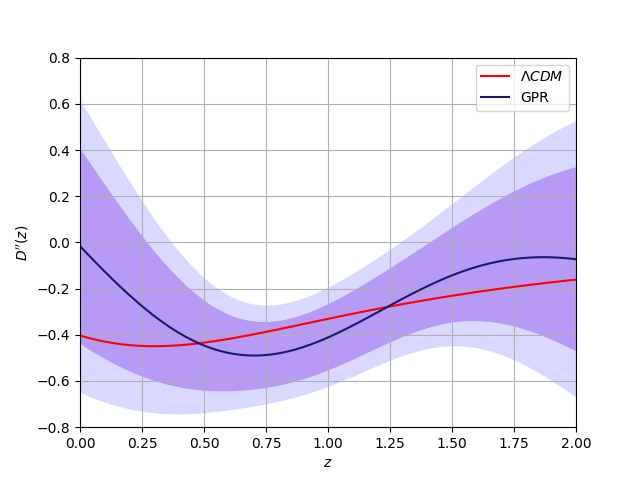}
    \caption{Reconstruction of the second derivative of the normalized comoving distance.}
    \label{Dpp}
\end{figure}

\begin{figure}[t!]
    \includegraphics[width=8cm, height=5.5cm]{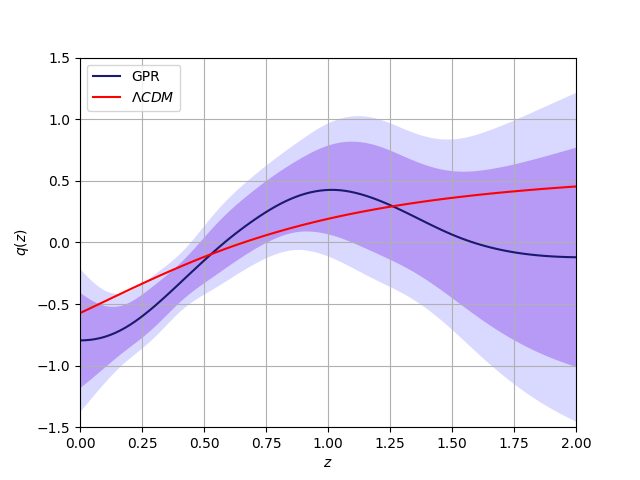}
    \caption{Reconstruction of the deceleration parameter.}
    \label{qmodel}
\end{figure}

From the same dataset of $D'(z)$ the derivative $D''(z)$ is reconstructed using the \texttt{GaPP} 
package as explained in Section \ref{sec:GPK}. On the other hand, Eq. (\ref{Dprime}) can be differentiated 
analytically and evaluated for the $\Lambda$CDM density parameters to obtain $D''(z)$. 
Figure \ref{Dpp} shows the predictions for $D''(z)$ and a comparison with $\Lambda$CDM. 
Finally, from Eq. (\ref{q}) and the GPR predictions of $D'(z)$ and $D''(z)$ a model of the 
deceleration parameter is produced as in the previous cases. The regression is compared 
with $\Lambda$CDM in Figure \ref{qmodel}. We see again some agreement between our 
reconstruction and the standard model, although an important thing to note is that there 
is a region where $\Lambda$CDM remains outside the 1$\sigma$ contour and it is really 
close to being outside the 2$\sigma$ one. This could indicate some actual evidence in favor 
of our reconstruction or at least highlight a tension existing within $\Lambda$CDM. 
Similar discrepancies have been noted in previous studies, where deviations from $\Lambda$CDM 
behavior were observed in various cosmological datasets \cite{Lin:2019cuy, Gomez-Valent:2018gvm, 
Haridasu:2018gqm, Mukherjee:2020vkx}. These findings suggest potential new physics beyond the 
standard cosmological model or the need for refined cosmological parameters which calls 
for further investigation.

\begin{figure}[t!]
    \includegraphics[width=8cm, height=5.5cm]{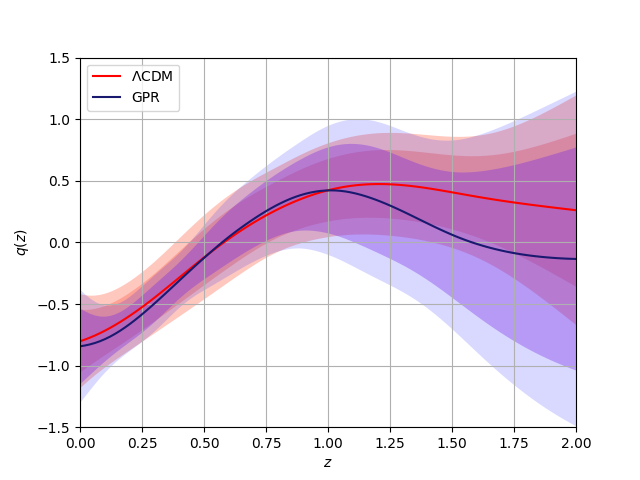}
    \caption{Comparison between the reconstruction of $q(z)$ and $\Lambda$CDM model using a mock data set.}
    \label{qmock}
\end{figure}

\subsection{Deceleration parameter reconstruction with a mock data set from $\Lambda$CDM}

If the observables are indeed produced by the $\Lambda$CDM model (that is to say that the 
standard model is the ``real'' one), a regression using artificial data that was 
produced by the $\Lambda$CDM model should be quite similar to the model obtained from 
the ``real'' data. To verify this, we produced a mock dataset of $H(z)$ around the values 
obtained by evaluating Eq. ($\ref{Hfriedmann}$) for the density parameters given 
by $\Lambda$CDM cosmology from Planck results \cite{aghanim2020planck}. 
Then, the whole procedure to obtain $q(z)$ was repeated but this time using the mock 
dataset so that a comparison with the previous reconstruction could be made. 
The result and comparison is shown in Figure $\ref{qmock}$. As expected, the mock 
data set regression model is into the $2\sigma$ confidence level of the reconstruction 
from the observations. This indicates that, even if the standard model does not 
reproduce the observables exactly or does it with some caveats, it can emulate the 
general observed behavior pretty well.

\subsection{Reconstruction of the jerk parameter}

\begin{figure}[t!]
    \includegraphics[width=8cm, height=5.5cm]{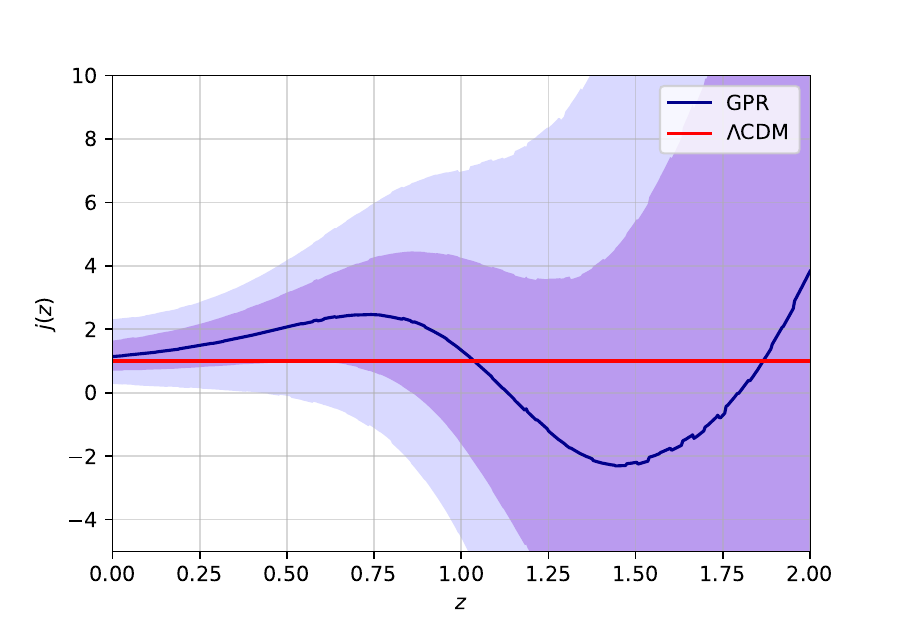}
    \caption{Comparison between the reconstruction of $j(z)$ and $\Lambda$CDM model.}
    \label{jmodel}
\end{figure}

From the same $D'(z)$ dataset, derived in Section \ref{sec:q}, one can further reconstruct 
the third derivative $D'''(z)$ along with the second derivative $D''(z)$ employing 
the \texttt{GaPP} package as explained in Section \ref{sec:GPK}. With these reconstructed functions, 
the evolution for the cosmological jerk parameter can be obtained from Eq. (\ref{eq:j}), 
as a function of the redshift $z$. This regression is compared with the $\Lambda$CDM 
case in Figure \ref{jmodel}. We find that our reconstruction includes the $\Lambda$CDM 
model (i.e., $j=1$) within the $1\sigma$ confidence level. The mean of the reconstructed 
function clearly indicates that $j$ has a non-monotonic evolution, which is in agreement 
with the previous findings~\cite{Mukherjee_2021}. 

\subsection{Using GPR as an interpolation in a Model-Independent way}

Throughout this work, we have demonstrated how a GPR can be utilized in a non-parametric 
manner to study cosmological quantities. However, there is an alternative approach to 
leverage the properties of a GPR which we would like to mention. This method also employs a Gaussian Process but in a 
model-independent manner, as it involves inferring parameter values using datasets and 
Bayesian statistics. By doing so, we can directly compare our model-independent 
reconstruction against the standard model using Bayesian evidence and maximum log-likelihood.


The GPR in this approach is done over ``nodes''. These nodes can vary in height (their 
ordinate position), and these ``variable heights'' work as the new parameters of the 
reconstruction \cite{AlbertoVazquez:2012ofj, Hee:2016nho}. For $n$ nodes we have $n$ variable heights and, as such, $n$ new parameters 
which need to be inferred. This method has been used before with the equation of 
state of Dark Energy \cite{gerardi2019reconstruction}, the interaction kernel of an 
IDE (interacting Dark Energy) model \cite{Escamilla:2023shf}, and the cosmic 
reionization history \cite{Krishak:2021fxp, Mukherjee:2024cfq}.

\section{Discussion and Conclusions}\label{sec:conclusions}

Although Gaussian Process Regression (GPR) does not yield an explicit form of the relationship 
between variables, it remains a robust method for making predictions given a particular 
set of observables. It reconstructs functions effectively without needing prior 
assumptions about their behavior, leveraging libraries like \texttt{GaPP} to predict higher derivatives, 
such as \(D''(z)\) and \(H'(z)\), which is particularly valuable in cosmological analyses.

GPR has been extensively applied in cosmology, spanning from reconstructing the dark energy 
equation of state $w(z)$ to cosmographical studies. This flexibility allows GPR to adapt 
to diverse datasets, making it a powerful tool for probing dark energy and other cosmological 
phenomena. In gravitational wave cosmology, GPR has been instrumental in reconstructing 
the luminosity distance from simulated data \cite{Shah:2023rqb}, enabling non-parametric 
inference of the Hubble parameter and forecasting deviations from the standard $\Lambda$CDM model. 
Additionally, GPR has been employed in large-scale structure studies, such as reconstructing 
the growth rate of cosmic structures $f\sigma_8(z)$ from redshift space distortions \cite{Calderon:2022cfj, Calderon:2023msm}. 
These applications highlight GPR's versatility in handling diverse cosmological datasets 
and addressing critical questions within the cosmological context.

However, GPR is constrained by the range of observed data, limiting its predictive accuracy 
outside this interval. Furthermore, uncertainties in derivative function reconstructions 
increase beyond the data range, which can impact the reliability of extrapolations. The 
choice of kernel function in GPR is pivotal, influencing prediction means and covariances 
significantly. Despite methods like cross-validation and Bayesian model selection to aid 
kernel selection, the optimal choice remains non-trivial, affecting the quality of reconstructions.

Furthermore, random number generation plays a subtle but crucial role in GPR applications. 
While tools like \texttt{scikit-learn} utilize robust generators like Permuted Congruential 
Generator 64-bit (PCG-64) \cite{oneill:pcg2014}, the quality of these generators in high-dimensional spaces, 
as highlighted in prior studies \cite{Luscher:1993dy, James:1993np, Shchur:1998wi}, warrants scrutiny. Though we found no immediate issues 
with the generators used in this work, but exploring alternatives like \texttt{RANLUX}\footnote{http://luscher.web.cern.ch/luscher/ranlux} \cite{James:1993np}, known for 
its high-quality randomness, could further ensure reliability. Such considerations are important 
in cosmological contexts, where multi-parameter reconstructions are common, and small biases 
can propagate into significant systematic errors \cite{God_owski_2012, Pajowska_2019}.

Comparing GPR with other parametric and non-parametric methods, principal component 
analysis \cite{LIU2019100379} (PCA), logarithmic parametrization \cite{Mamon_2017}, rational 
parametrization \cite{physics4040090}, Bayesian methods \cite{xu}, reveals trade-offs 
between flexibility and interpretability. While PCA simplifies data dimensionality effectively, 
it may overlook intricate data complexities that GPR can capture. Bayesian methods provide 
comprehensive probabilistic frameworks but often require detailed prior information 
and intensive computational resources.
\\

In summary, Gaussian Processes offer a powerful and flexible tool for cosmological analyses, 
enabling model-independent reconstructions and effective uncertainty handling. Despite 
computational challenges and kernel sensitivity, their widespread application in cosmology 
demonstrates their potential to provide nuanced insights into the evolutionary history 
of our Universe.

\section*{Acknowledgments}
PM acknowledges funding from Anusandhan National Research Foundation (ANRF), Govt of India under the National PostDoctoral Fellowship (File no. PDF/2023/001986). JJV, LAE and JAV acknowledge the support provided by FORDECYT-PRONACES-CONACYT/304001/2020 and UNAM-DGAPA-PAPIIT IN117723.

\appendix
\section*{Appendix A: GPR figures}

\begin{figure*}
    \centering
    \includegraphics[width=15cm,height=10cm]{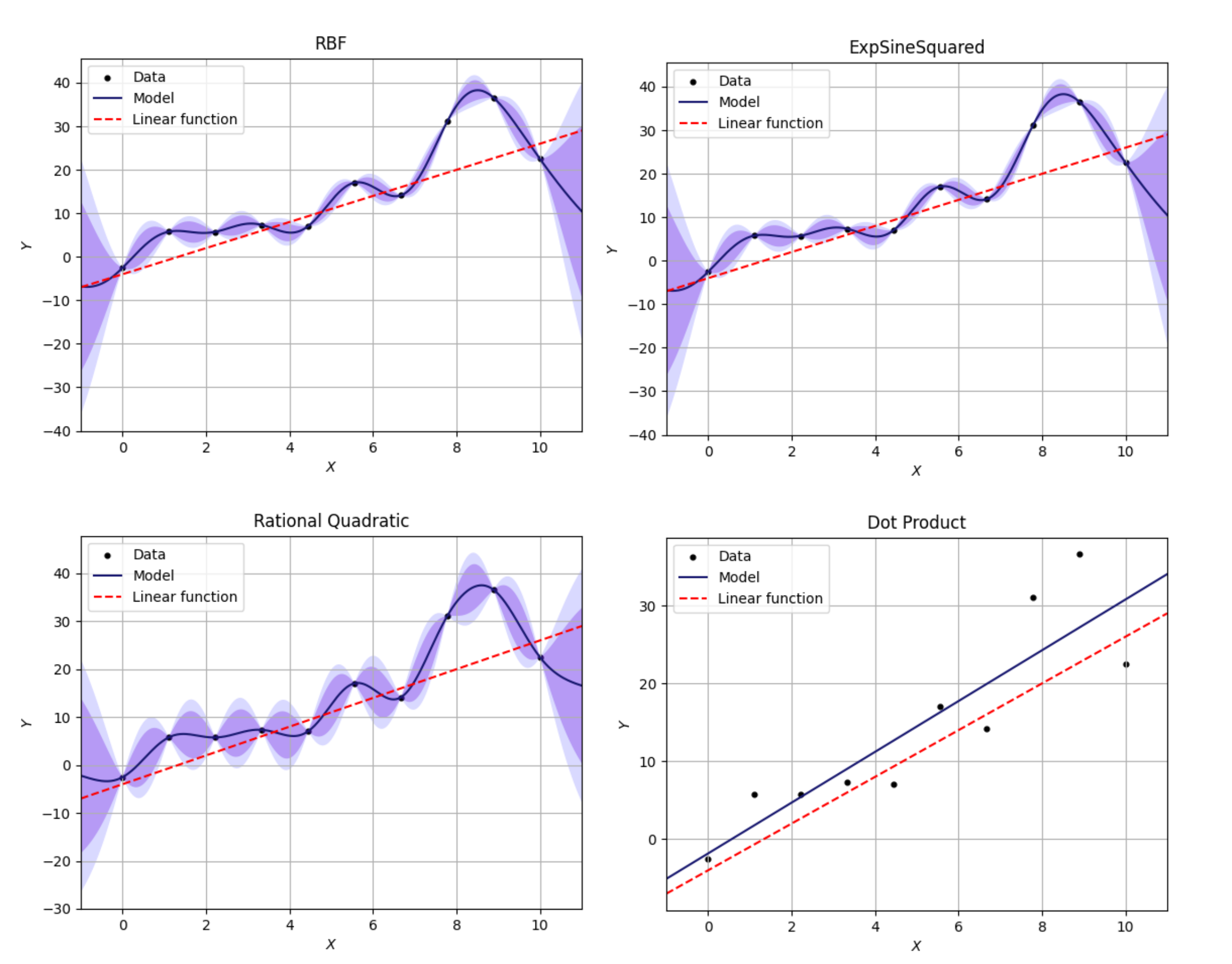}
    \caption{Models obtained from different kernels for the same linear regression with null variances.}
    \label{KernelsNullLineal}
\end{figure*}
\begin{figure*}
    \centering
    \includegraphics[width=15cm,height=10cm]{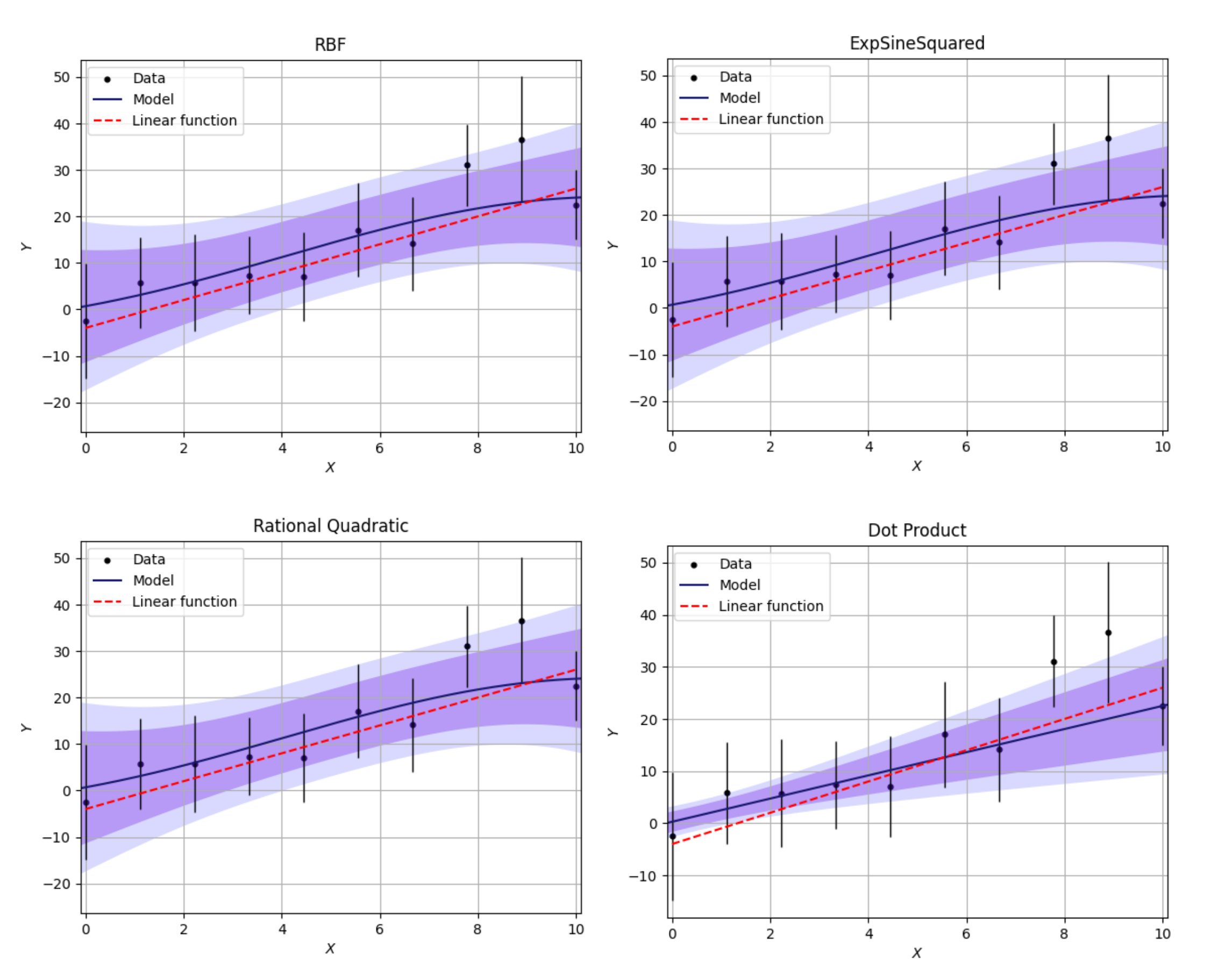}
    \caption{Linear regression models from different kernels with non-zero variances.}
    \label{KernelsNoNullLinear}
\end{figure*}
\begin{figure*}
    \centering
    \includegraphics[width=15cm,height=10cm]{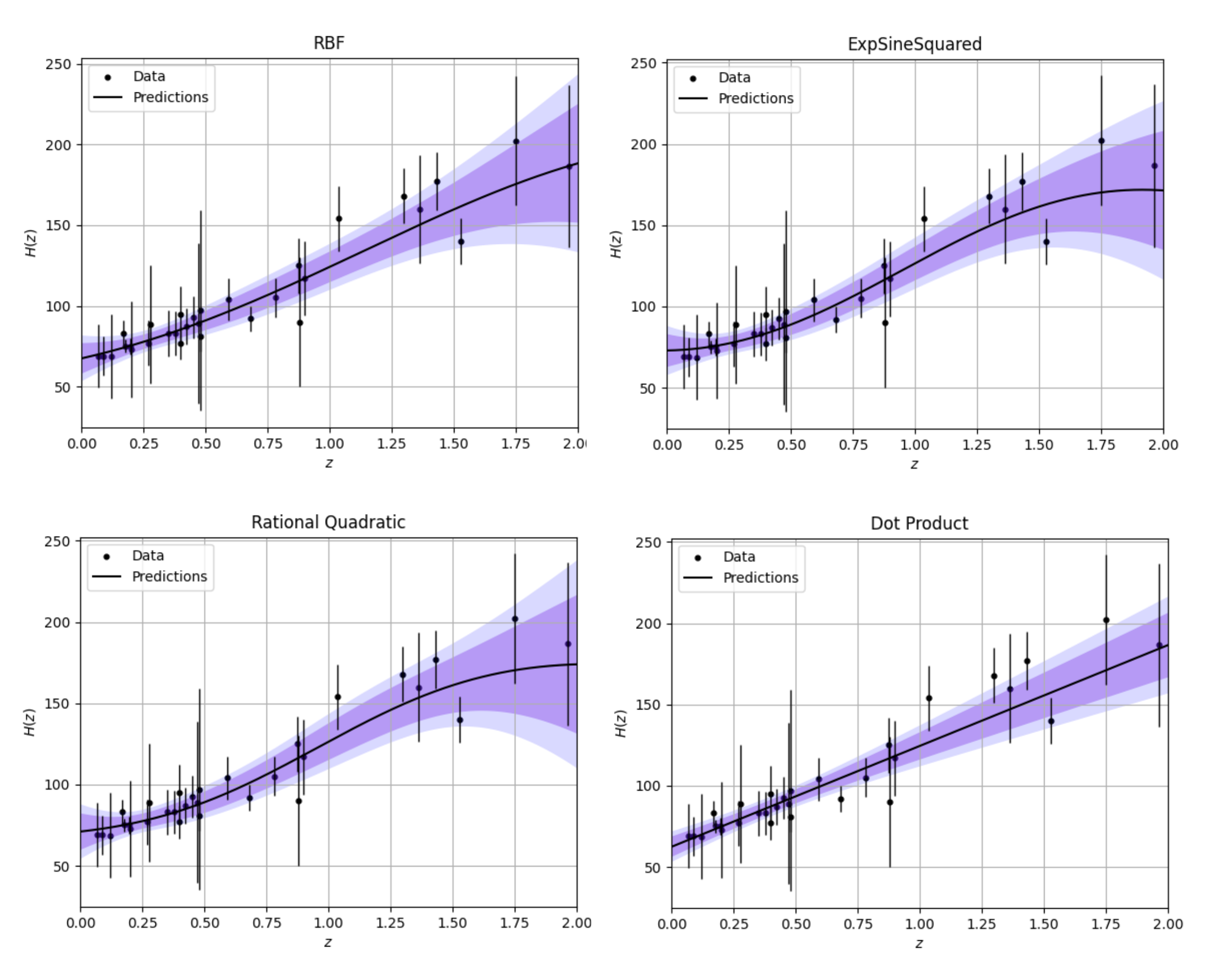}
    \caption{Regression models of Hubble Parameter for different kernels.}
    \label{Hubblekernels}
\end{figure*}
\begin{figure*}
    \centering
    \includegraphics[width=15cm,height=10cm]{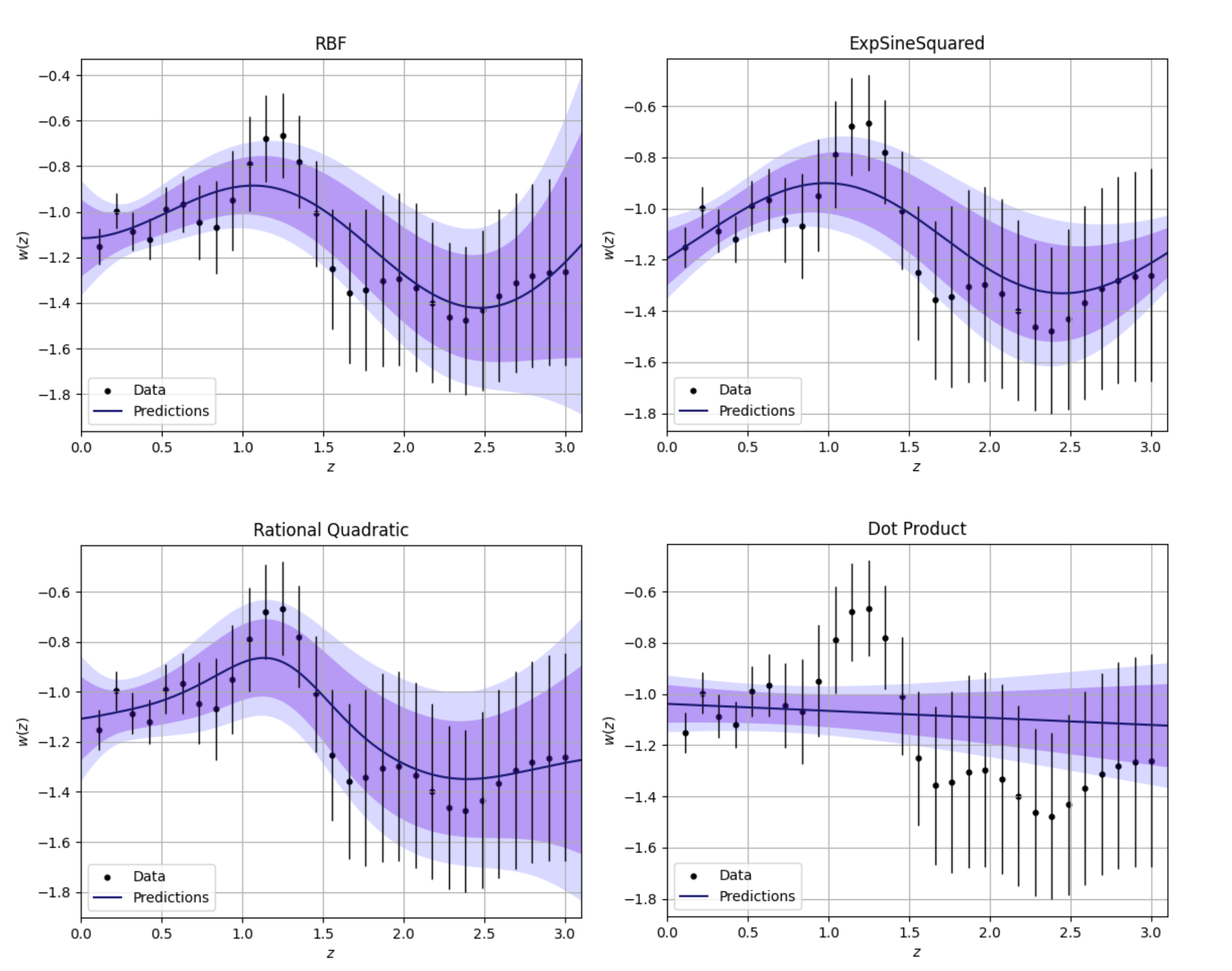}
    \caption{Regression models of Dark Energy for different kernels.}
    \label{DEkernels}
\end{figure*}

\bibliography{bibfile}

\end{document}